\documentclass[10pt,aps,prb,twocolumn,showpacs,amsmath,amssymb,superscriptaddress,bibnotes]{revtex4-2}

\usepackage{graphicx}
\usepackage{color}
\usepackage{braket}
\usepackage{amsfonts}

\usepackage{soul}
\usepackage{siunitx}

\usepackage{mathrsfs}
\usepackage{bbm}
\usepackage[dvipsnames]{xcolor}
\usepackage{appendix}
\usepackage{verbatim}
\usepackage{bbold}

\usepackage[normalem]{ulem}

\usepackage[unicode]{hyperref} 
\usepackage{blkarray} 

\newcommand{\m}[1]{\mathrm{#1}}

\begin{document}
    
	\title{Parity-Resolved Quantum Capacitance and Quantum Inductance in Topological, Trivial, and Normal Nanowire Interferometers}

	\author{Viktoriia Pinchenkova}
	\affiliation{Department of Physics, University of Basel, Klingelbergstrasse 82, CH-4056 Basel, Switzerland}
    \author{Valerii K. Kozin}
	\affiliation{Department of Physics, University of Basel, Klingelbergstrasse 82, CH-4056 Basel, Switzerland}
    \author{Maximilian Hünenberger}
	\affiliation{Department of Physics, University of Basel, Klingelbergstrasse 82, CH-4056 Basel, Switzerland}
    \author{Daniel Loss} 
    \affiliation{Department of Physics, University of Basel, Klingelbergstrasse 82, CH-4056 Basel, Switzerland}
\affiliation{Physics Department, King Fahd University of Petroleum and Minerals, 31261, Dhahran, Saudi Arabia}
\affiliation{Quantum Center, KFUPM, Dhahran, Saudi Arabia}
\affiliation{RDIA Chair in Quantum Computing}
    \author{Jelena Klinovaja}
	\affiliation{Department of Physics, University of Basel, Klingelbergstrasse 82, CH-4056 Basel, Switzerland}
	
	\date{\today}
	
\begin{abstract}

Quantum-capacitance measurements convert the curvature of a quantum-dot energy in a flux-threaded nanowire loop into fast parity-sensitive signals and, therefore, have become a promising readout tool for Majorana devices.
However, Majorana-like quantum-capacitance responses can also arise from topologically trivial Andreev bound states, making capacitance alone insufficient to identify a topological phase.
To analyze this problem, we consider a quantum dot coupled to both ends of four nanowire realizations:
    a topological nanowire hosting Majorana bound states,
    non-topological superconducting nanowires hosting one or two Andreev bound states,
    and a fully normal nanowire.
Motivated by proposals to use quantum inductance as an additional phase-sensitive probe, we compute both the parity-resolved quantum capacitance $C_\m{Q}$ and inverse quantum inductance $L_\m{Q}^{-1}$ as functions of the magnetic flux by exact diagonalization.
We show that signatures associated with zero-energy Majorana bound states, such as $h/e$ periodicity and an $h/(2e)$ flux shift between even and odd parity sectors in $C_\mathrm{Q}$ and $L_\mathrm{Q}^{-1}$, are not sufficient indicators of topological superconductivity.
In certain realistic parameter regimes, similar Majorana-like behavior can arise from a trivial Andreev bound state and even from a purely normal nanowire.
By contrast, two nearly zero-energy Andreev bound states can generate a pronounced $h/(2 e)$-periodic component associated with charge-$2e$ transfer, providing a clear non-Majorana signature.
A low-energy projection shows that the Majorana, single-Andreev-state, and normal cases can be mapped onto the same minimal low-energy model explaining their similar flux-dependent responses despite their different physical origins. 

\end{abstract}
	
\maketitle

\section{Introduction}\label{sec:Introduction}

Majorana bound states (MBSs) in one-dimensional topological superconductors provide a nonlocal fermionic degree of freedom and are a central ingredient in proposals for topological quantum computation~\cite{Kitaev2001,Lutchyn2010,Oreg2010,Alicea2012,LaubscherKlinovaja2021,Bonderson2008,Karzig2017}. Semiconductor-superconductor nanowires have produced several experimental signatures consistent with Majorana physics, including a zero-bias peak, quantum-dot spectroscopy of near-zero modes, and island experiments showing exponentially small mode splitting~\cite{Mourik2012,Deng2016,Albrecht2016}. However, these probes are primarily sensitive to low-energy states rather than directly to topology. In realistic finite and inhomogeneous nanowires, trivial Andreev bound states (ABSs) and quasi-Majorana modes can remain near zero energy and mimic key Majorana signatures~\cite{Kells2012,Lee2012,Rainis2013,Cayao2015,Ptok2017,Liu2017,Reeg2018_ABS,Peñaranda2018,Moore2018,Sauls2018,Vuik2019,StanescuTewari2019,Woods2019_ABS,Liu2019,Chen2019,Awoga2019,Pan2020,Prada2020,Alspaugh2020,Jünger2020,Hess2021,Valentini2021,Hess2023,ArayaDay2025}. 
This makes it essential to test which proposed readout signatures are uniquely topological and which can be reproduced by conventional subgap physics.

Quantum-capacitance measurements have recently emerged as a promising parity-sensitive probe. In a flux-threaded nanowire--quantum-dot loop, the quantum capacitance is set by the curvature of the many-body energy with respect to the dot potential, similarly to dispersive gate-sensing and mesoscopic-capacitor measurements~\cite{Buttiker1993,Colless2013}. Recent interferometric parity measurements in semiconductor-superconductor hybrid devices demonstrated the experimental potential of this approach~\cite{1Microsoft,3Microsoft,Majorana2}. Theoretical and numerical studies connected the capacitance response to Rabi-oscillation measurements, large-scale dynamical simulations, and the assessment of MBS quality~\cite{Sankar2025,2Microsoft,Dourado2026}. At the same time, analyses of overlapping MBSs with a near-zero energy splitting and quasi-Majorana modes showed that parity-dependent capacitance oscillations are not unique to well-separated topological MBSs~\cite{Stanescu2026}. The inverse quantum inductance has likewise been proposed as a complementary phase-sensitive diagnostic of fermion-parity switching in Majorana nanowires~\cite{Tewari2026}. 

These developments motivate the central question addressed here: Are the characteristic $h/e$-periodic quantum-capacitance and quantum-inductance oscillations, together with a half-period flux shift between two parity sectors, specific to topological MBSs, or can similar signatures also arise in non-topological superconducting and even fully normal interferometers?
We answer this question in a unified microscopic model of a quantum dot (QD) coupled to both ends of a nanowire interferometer threaded by magnetic flux. We compare four realizations on equal footing: a topological superconducting nanowire hosting MBSs, non-topological superconducting nanowires hosting either one or two ABSs, and a fully normal nanowire. For all cases, we compute the flux dependence of the quantum capacitance $C_\m{Q}$ and inverse quantum inductance $L_\m{Q}^{-1}$ from the exact many-body energy curvatures. 

Our main result is that the standard Majorana-like pattern is not unique to the topological phase and can even appear in a fully normal, nonsuperconducting interferometer.
While the MBSs with nearly zero energy display the expected $2\pi$-periodic response in $\phi=2\pi\Phi/(h/e)$ (with $\Phi$ being the magnetic flux), and a $\pi$ shift between the two fermion-parity sectors, a single delocalized ABS can produce nearly identical traces in certain realistic parameter regimes.
Furthermore, a normal nanowire interferometer can mimic the same flux dependence when adjacent particle-number sectors are compared.
The common origin of these false positives is a low-energy two-level interference structure that emerges after projecting the MBS, single-ABS, and normal systems onto their relevant low-energy subspaces.
Importantly, neither quantum capacitance, nor quantum inductance, nor their combination can serve as a smoking gun evidence for MBSs as our counterexamples demonstrate.
Thus, capacitance and inductance interferometry are powerful probes of coherent end-to-end coupling and parity-dependent dynamics, but $h/e$ periodicity and a half-period sector shift are not sufficient indicators of topological superconductivity.

The two-ABS case provides an important contrast.
When two nearly zero-energy ABSs participate, charge-$2e$ processes generate a pronounced $\pi$-periodic component in both $C_\m{Q}(\phi)$ and $L_\m{Q}^{-1}(\phi)$, giving a clear non-Majorana signature.
However, if the energy of one of the ABSs is tuned closer to the bulk-gap edge and effectively mimics a bulk state, the system crosses over to a single-ABS-like regime and can again display Majorana-like traces.

The paper is organized as follows. In Sec.~\ref{sec:model}, we introduce the model of four interferometric setups and define the quantum capacitance and inverse quantum inductance. In Sec.~\ref{sec:topological}, we analyze the topological superconducting interferometer hosting MBSs and derive a minimal low-energy model that explains the $2\pi$-periodic response and the $\pi$ phase shift between the two parity branches. In Sec.~\ref{sec:non-topological}, we turn to non-topological superconducting interferometers with one or two ABSs and show when their responses mimic, or differ from, the MBS case. In Sec.~\ref{sec:normal}, we consider a fully normal nanowire interferometer and demonstrate that similar flux-dependent traces can arise even without superconductivity.
Finally, we conclude in Sec.~\ref{sec:conclusion}.

\section{Model of parity-resolving interferometers}\label{sec:model} 

\subsection{Nanowire}

\begin{figure}[tb]
    \centering   \includegraphics[width=0.48\textwidth]{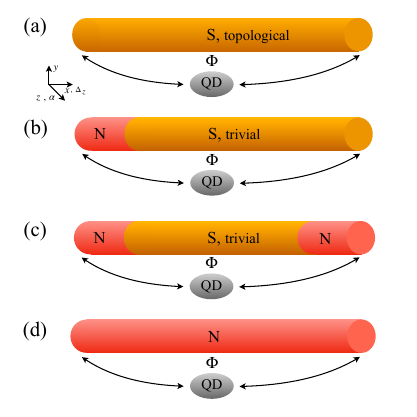}
    \caption{Different interferometric setups considered in this work. A Rashba nanowire (NW) is aligned along the $x$ axis. The SOI vector points in the $z$ direction and the magnetic field in the $x$ direction. A quantum dot (gray) is coupled to the left and right ends of the NW, forming an interferometric loop that encloses a magnetic flux $\Phi$. (a) The NW is covered by a superconductor, which induces superconductivity (S) via the proximity effect. The NW is tuned to the topological regime, where it can host MBSs localized at its ends. (b) The NW is tuned to the trivial phase. The left segment is not covered by the superconductor and remains normal (N), allowing for an ABS localized near the S–N boundary. (c) Both left and right segments are normal, such that the non-topological NW can host two ABSs localized near the corresponding S–N boundaries. (d) The superconductor is removed, and the entire NW is in a normal state.
    }
    \label{fig:model}
\end{figure}

We start by considering a one-dimensional Rashba nanowire aligned along the $x$ direction, as shown in Fig.~\ref{fig:model}. The nanowire is coupled to a parent superconductor, which induces a position-dependent superconducting gap $\Delta(x)$. The Rashba spin-orbit interaction (SOI) vector with strength $\alpha(x)$ is aligned along the $z$ direction and sets the spin quantization axis. A magnetic field with the Zeeman energy $\Delta_\m{Z}(x)$ is applied along the $x$ axis. In terms of the electron creation and annihilation operators $c_{ n , \sigma }^\dagger$ and $c _ { n , \sigma }$, where $\sigma = \uparrow, \downarrow$ denotes the spin and $n$ labels the lattice site, the nanowire Hamiltonian can be written as 
\begin{align}
H_\mathrm{NW} = & \sum_{n=1}^N \, \sum _{ \sigma, \sigma^{ \prime } } \, c _ { n , \sigma } ^ { \dagger } [ (t_{n+\frac{1}{2}}+t_{n-\frac{1}{2}}-\mu_n ) \delta_{\sigma \sigma^{\prime}} \nonumber \\
& +\Delta_{\m{Z}, n} \sigma_{\sigma \sigma^{\prime}}^x ]c_{n, \sigma^{\prime}} \nonumber \\
& -\left\{ c_{n, \sigma}^{\dagger}\left[t_{n+\frac{1}{2}} \delta_{\sigma \sigma^{\prime}}-i \alpha_{n+\frac{1}{2}} \sigma_{\sigma \sigma^{\prime}}^z\right] c_{n+1, \sigma^{\prime}}  \right. \nonumber \\
& +\Delta_n c_{n, \downarrow}^{\dagger} c_{n, \uparrow}^{\dagger}+\text { H.c.} \left. \right \},
\end{align}
where $N$ is the total number of sites, $t_n$ is the nearest-neighbor tunneling amplitude, and $\mu_n$ is the chemical potential \cite{Hess2021}. Furthermore, $\delta_{\sigma \sigma'}$ denotes the Kronecker delta, and $\sigma^{i}$ with $i \in \{x,y,z\}$ denotes Pauli matrices representing the electron spin. We are interested in four different setups that are presented in Fig.~\ref{fig:model}. 

\emph{Topological nanowire.} This case is shown in Fig.~\ref{fig:model}(a) and characterized by uniform parameters along the nanowire, namely, $t_n = t$, $\mu_n = \mu$, $\Delta_{n} = \Delta$, $\alpha_n = \alpha$, and $\Delta_{\m{Z},n} = \Delta_\m{Z}$. The nanowire is tuned to the topological phase, determined by the condition $|\Delta_\m{Z}| > \sqrt{\mu^2 + \Delta^2}$, where it can host two MBSs localized at its ends. Note that the magnetic field suppresses the bulk gap of the parent superconductor, resulting in a decrease of the proximity-induced superconducting gap with increasing Zeeman energy:
\begin{equation}\label{eq:Delta}
\Delta=\Delta_0 \sqrt{1-\left(\Delta_\m{Z} / \Delta_\m{Z}^c\right)^2},
\end{equation}
where $\Delta_0$ is the maximal value, and the superconducting gap vanishes at the critical field $\Delta_\m{Z}^c$. In the rest of the paper, we use the value $\Delta_\m{Z}^c = 1.75$ meV.

\emph{Non-topological nanowire}. We now assume that one or two nanowire segments are not proximitized by the superconductor and therefore remain in the normal state, as shown in Figs.~\ref{fig:model}(b) and~\ref{fig:model}(c). The left (right) normal section consists of $N_1$ ($N_2$) sites, while the superconducting section contains $N_s$ sites, so that the total number of sites is given by $N = N_1 + N_s + N_2$. To describe nonuniform system parameters, we first define the boundary between the left normal section and the superconducting section as $N_l = N_1+1/2$, and the boundary between the right normal section and the superconducting section as $N_r = N_1+N_s + 1/2$. We set the superconducting term to be nonzero only in the superconducting region:
\begin{equation}
\Delta_n = \Delta\left[\theta\left(n-N_l\right)-\theta\left(n-N_r\right)\right],
\end{equation}
with $\theta(n)$ being the Heaviside step function with $\theta(0) = 1/2$, and $\Delta$ determined by Eq.~(\ref{eq:Delta}). To ensure that the system remains in the non-topological phase, we set the Zeeman energy and Rashba SOI to zero in the superconducting section:
\begin{equation}
\Delta_{\m{Z}, n}  =\Delta_\m{Z} [\theta\left(N_l-n\right) + \theta\left(n-N_r \right)], 
\end{equation}
\begin{equation}
\alpha_n  =\alpha \, [\theta\left(N_l-n\right) + \theta\left(n-N_r\right)],
\end{equation}
with finite Zeeman energy $\Delta_\m{Z}$ and SOI strength $\alpha$ in the $N_1$ and $N_2$ sections.

The superconducting and normal sections are characterized by different nearest-neighbor tunneling amplitudes $t_s$ and $t$, respectively:
\begin{align}
t_n= & t \, [\theta\left(N_l-n\right) + \theta\left(n-N_r\right)] \nonumber \\  &+  t_s\left[\theta\left(n-N_l\right)-\theta\left(n-N_r\right)\right].
\end{align}
This difference originates from the renormalization of the mass inside the superconducting section resulting from the metallization effects induced by the thin superconducting shell~\cite{Reeg2017,Reeg2018, Reeg2018_PRB,Winkler2019, Woods2019,Kiendl2019}. Finally, we put the chemical potential to be equal to $\mu$ in the normal sections and $\mu_s$ in the superconducting section: 
\begin{align}
\mu_n= & \mu \, [\theta\left(N_l-n\right)+\theta\left(n-N_r\right)] \nonumber \\
&  +\mu_s\left[\theta\left(n-N_l\right)-\theta\left(n-N_r\right)\right].
\end{align}
In these two setups, ABSs can form close to the boundaries between the superconducting and normal sections. We choose the parameter regime, where the level spacing between ABSs, $2 \alpha/N_i$ with $i = 1,2$, is larger than the superconducting gap $\Delta_0$, such that the system hosts only a single ABS per boundary~\cite{Hess2021}.

\emph{Nonsuperconducting nanowire}.  In our final model, presented in Fig.~\ref{fig:model}(d), we set the superconducting gap to zero, $\Delta_{n} = 0$, and focus only on the nanowire in the normal state. We also remove SOI $\alpha_n = 0$, because it does not influence the result, as will be discussed below in Sec.~\ref{sec:normal}. The remaining parameters are taken to be uniform along the nanowire: $t_n = t$, $\mu_n = \mu$, and $\Delta_{\m{Z},n} = \Delta_\m{Z}$.

\subsection{Coupling to the quantum dot}

All four nanowires, described in the previous subsection, are tunably coupled to a QD at the left and right ends, forming a closed loop. The out-of-plane magnetic field creates a magnetic flux $\Phi$ through this interferometric loop. We consider a nonsuperconducting QD and treat it as the zeroth site of our tight-binding model with the energy $V_\m{QD}$. We also add the Zeeman energy $\Delta_\m{Z}$ of the magnetic field to remove the spin degeneracy. The Hamiltonian of the dot is thus given by
\begin{align}
& H_{0}  =  \sum _ { \sigma , \sigma ^ { \prime } }  c _ { 0 , \sigma }^{ \dagger }  \left(V_\m{QD} \delta_{\sigma \sigma^{\prime}} + \Delta_\m{Z} \sigma_{\sigma \sigma^{\prime}}^x\right) c_{0, \sigma^{\prime}},
\end{align}
with eigenenergies $V_\m{QD} \pm \Delta_\m{Z}$. The dot is tunably coupled to the left and right ends of a nanowire via normal and SOI tunneling processes. The strength of the tunnelings is controlled by the dimensionless parameters $\lambda_\m{L}$ and $\lambda_\m{R}$ ($\lambda_\m{L}=\lambda_\m{R}\equiv\lambda_\m{QD}$ unless stated otherwise), and the corresponding Hamiltonian terms are
\begin{align}
& H_\m{L}  =  \lambda_\m{L} e^{i \phi }\sum_{\sigma, \sigma^{\prime}} c_{0, \sigma}^{\dagger}  [ -t  \delta_{\sigma \sigma^{\prime}} + i \alpha   \sigma^z_{\sigma \sigma^{\prime}}] c_{1, \sigma^{\prime}} + \textup{H.c.,}\\
&  H_\m{R} =  \lambda_\m{R} \sum_{\sigma, \sigma^{\prime}} c_{0, \sigma}^{\dagger}  [ -t  \delta_{\sigma \sigma^{\prime}} + i \alpha   \sigma^z_{\sigma \sigma^{\prime}}] c_{N, \sigma^{\prime}} + \textup{H.c.}
\end{align}
The phase $\phi$ is the phase difference between the right and left tunnelings controlled by the magnetic flux as $\phi = 2 \pi \Phi/\Phi_0$ (Peierls phase), with the magnetic flux quantum $\Phi_0 = h/e$. We consider $H_\mathrm{L}$ and $H_\mathrm{R}$ to be weak links ($\lambda_\m{L}, \lambda_\m{R} \ll 1$).
Therefore, we neglect superconducting phase gradients inside the nanowire 
\cite{soninTheoryPlanarBallistic2026}.
Additionally, we chose a gauge such that the phase is present only on the left side. The total Hamiltonian of our model is given by the sum of all terms described above:
\begin{equation}\label{eq:Ham_tot}
    H_\m{tot} = H_\m{NW} + H_0 + H_\m{L} + H_\m{R}.
\end{equation}
We assume that the contributions from a readout system and all sources of noise are negligible. Throughout, we work at zero temperature and model
the QD as a noninteracting single orbital~\cite{Sankar2025}.

In the next sections, we investigate the behavior of the quantum capacitance $C_\m{Q}$ and the inverse quantum inductance $L_\m{Q}^{-1}$ in dependence on the magnetic flux $\Phi$,  or equivalently the phase difference $\phi \propto \Phi$, for the given interferometers. The quantum capacitance of our systems is determined as a variation of a charge on the QD with respect to a variation of the potential $V_\m{QD}$, or equivalently as
\begin{equation}\label{eq:quantum_capac}
C_\m{Q} = - e^2 \tilde{\alpha}^2 \frac{\partial^2 E}{\partial V_\m{QD}^2},
\end{equation}
where we put the lever arm $\tilde{\alpha}=1$, and $e^2=0.16\,$fF$\cdot$meV. The inverse quantum inductance is determined as
\begin{equation}\label{eq:quantum_induc}
L_\m{Q}^{-1} = \left ( \frac{e}{\hbar} \right )^2\frac{\partial^2 E}{\partial \phi^2},
\end{equation}
where $(e /\hbar)^2 = 0.37 \cdot10^{3}\,$[$\mu$H$\cdot$meV]$^{-1}$. For brevity, in the rest of the paper we refer to $L_\m{Q}^{-1}$ simply as the quantum inductance. 
We define the even/odd many-body energies using the Bogoliubov--de Gennes (BdG) ground state and the lowest quasiparticle excitation. Namely, in one parity sector, we take $E$ to be the ground state energy $E_\m{GS} = \frac{1}{2}\sum_{E_n <0} E_n$, where $E_n$ are the eigenvalues of $H_\m{tot}$ written in the BdG form. In another parity sector, we use the excited state energy $E = E_\m{GS} + E_+$, where $E_+$ is the lowest positive BdG eigenenergy. We then determine the parity of the ground state by calculating the sign of the Pfaffian of $-i$ times the BdG Hamiltonian written in the canonical Majorana basis~\cite{Kitaev2001}. When the Pfaffian is positive, the ground state is even, whereas an odd ground state yields a negative Pfaffian.

\begin{figure}[tb]
    \centering   \includegraphics[width=0.48\textwidth]{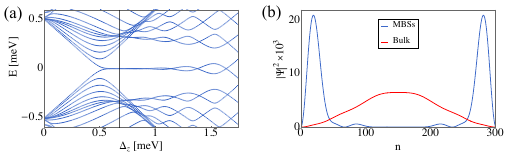}
    \caption{(a) Energy spectrum of the nanowire uncoupled from the QD in the topological interferometer, as a function of the Zeeman energy $\Delta_\m{Z}$. (b) Probability density of the MBSs and the lowest-energy bulk state at $\Delta_\m{Z} = 0.675\,$meV [indicated by the black line in (a)]. The chosen $\Delta_\m{Z}$ lies in the topological phase and yields nearly zero MBS energy splitting, resulting in states localized at the nanowire ends with negligible overlap. The parameters are listed in Table~\ref{tab:Majorana} in Appendix~\ref{app:param}. }
    \label{fig:Majorana}
\end{figure}

\section{Topological interferometer}\label{sec:topological}

In this section, we consider the interferometer shown in Fig.~\ref{fig:model}(a), where the nanowire is tuned to the topological phase and hosts MBSs localized at its ends. We begin our analysis by considering the nanowire decoupled from the QD. Figure~\ref{fig:Majorana}(a) shows the energy spectrum of the nanowire as a function of the Zeeman energy $\Delta_\m{Z}$. In Fig.~\ref{fig:Majorana}(b), we plot the probability density $|\Psi|^2$ of the MBSs and the lowest-energy bulk state at the chosen value of $\Delta_\m{Z} = 0.675\,$meV. We choose $\Delta_\m{Z}$ within the topological phase and such that the MBS energy splitting is close to zero. As a result, the MBSs are localized at the ends of the nanowire and have negligible overlap.

\begin{figure}[!bt]
    \centering   \includegraphics[width=0.48\textwidth]{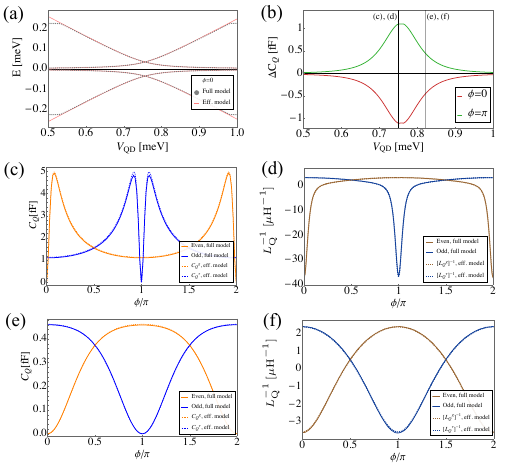}
    \caption{(a) Low-energy spectrum of the topological interferometer. (b) Difference between quantum capacitance in the even and odd parity sectors $\Delta C_\m{Q}$ as a function of $V_\m{QD}$. (c), (e) Even and odd branches of the quantum capacitance $C_\m{Q}$ and (d), (f) the quantum inductance $L_\m{Q}^{-1}$ at $V_\m{QD} = 0.75\,$meV and $V_\m{QD} = 0.82\,$meV [indicated by the black lines in (b)]. In all presented cases, $C_\m{Q}(\phi)$ and $L_\m{Q}^{-1}(\phi)$ exhibit a $2\pi$-periodicity and $\pi$ phase shift. The effective model [dashed lines in (c)-(f)] reproduces the results of the full Hamiltonian (solid lines). The parameters are listed in Table~\ref{tab:Majorana} in Appendix~\ref{app:param}.}
    \label{fig:Majorana_results}
\end{figure}

Next, we turn on the coupling between the nanowire and QD. Since we are interested in the weak-coupling regime, we set $\lambda_\m{QD} = 0.02$. The quantum capacitance is expected to be maximal for $V_\mathrm{QD}$ values in the vicinity of the avoided crossing between the MBS and QD energy levels [see Fig.~\ref{fig:Majorana_results}(a)]. In this range, Fig.~\ref{fig:Majorana_results}(b) shows the difference between quantum capacitance in the even and odd parity sectors $\Delta C_\mathrm{Q}$ for fixed $\phi = 0$ and $\pi$. To explore the dependence of $C_\m{Q}$ on the phase difference $\phi$, we choose two 
reference values of $V_\m{QD}$: one at resonance, $V_\m{QD} = 0.75\,$meV, and another slightly away from resonance, $V_\m{QD} = 0.82\,$meV.
We then plot the quantum capacitance and quantum inductance in the even and odd parity sectors as functions of $\phi$.
Figures~\ref{fig:Majorana_results}(c)--\ref{fig:Majorana_results}(f) reveal the main features of $C_\m{Q}(\phi)$ and $L_\m{Q}^{-1}(\phi)$ in the topological interferometric setup, namely a $2 \pi$ periodicity and a relative $\pi$ phase shift between the even and odd sectors, which agrees with the results obtained in the previous studies~\cite{1Microsoft,Sankar2025, Stanescu2026,Tewari2026}.

Our numerical results can be explained by a simplified effective model that takes into account only the low-energy subspace.
Since the bulk states and one level on the QD are well separated from zero energy, we can consider only MBSs in the nanowire with the energy splitting $2E_\m{M}$ and one level on the dot with the energy $E_\m{QD}$.
Both $E_\m{M}$ and $E_\m{QD}$ are assumed to be small compared to the superconducting bulk gap of the nanowire. 
The QD state couples to the MBS pair through the left and right ends of the nanowire via normal tunneling processes with amplitudes $g_\m{L}$, $g_\m{R}$ and Andreev tunneling processes with amplitudes $\tau_\m{L}$, $\tau_\m{R}$.
The effective low-energy model Hamiltonian is thus given by
\begin{align} \label{eq:Ham_eff}
H_\m{eff} & = E_{\mathrm{QD}} d^{\dagger} d  +2 E_{\mathrm{M}}\left(c^{\dagger} c-\frac{1}{2}\right) \nonumber \\ & + d^{\dagger} \left(g_{\mathrm{R}}c  +  \tau_{\mathrm{R}} c^{\dagger} \right)  + i d^{\dagger}  ( \tau_{\mathrm{L}}c^{\dagger} - g_{\mathrm{L}} c) + \textup{H.c.},
\end{align}
where $c^\dagger$ and $c$ create and annihilate an electron in the nanowire, while $d^\dagger$ and $d$ create and annihilate an electron in the QD. We choose a gauge such that the phase is present only on the left side, namely $g(\phi) = g_R - i g_L \equiv |g_R| - i |g_L| e^{-i(\phi + \phi_g)} $, $\tau(\phi) = \tau_R + i \tau_L \equiv |\tau_R| + i |\tau_L| e^{-i(\phi + \phi_\tau)}$, with $\phi$ being the flux-dependent phase difference between
right and left tunnelings. The phase offsets  $\phi_g$ and $\phi_\tau$ are flux-independent contributions. In the perturbative regime, these phases and amplitudes can, in principle, be obtained analytically \cite{QD_1}; however, the corresponding calculations are rather involved. For the analysis presented below, we therefore determine the parameters of the effective low-energy model by evaluating the perturbative expansion numerically.

The effective-model Hamiltonian is block-diagonal in the Fock space $|x y \rangle = |x \rangle_\m{NW} |y \rangle_\m{QD}$ with $x,y \in \{ 0,1\}$. The matrix of the first block can be written in the basis $\{|00\rangle, |11\rangle\}$ as
\begin{equation}\label{eq:Htau}
H^\tau = 
 \begin{pmatrix}
-E_\m{M} & -\tau^*(\phi) \\
 -\tau(\phi) &  E_\m{M} + E_\m{QD}\\
\end{pmatrix},
\end{equation}
while the matrix of the second block in the basis $\{|10\rangle, |01\rangle\}$ takes the form
\begin{equation}\label{eq:Hg}
H^g = 
 \begin{pmatrix}
E_\m{M} & g^*(\phi) \\
 g(\phi) &  -E_\m{M} + E_\m{QD}\\
\end{pmatrix}.
\end{equation}
Finding the lowest energies of each block and substituting them  into Eq.~(\ref{eq:quantum_capac}) for the quantum capacitance, we find
\begin{equation}\label{eq:CQ_eff1}
C_{\mathrm{Q}}^{\tau}(\phi)=\frac{2 e^2 \tilde{\alpha}^2\left|\tau(\phi)\right|^2}{[\left(E_{\mathrm{QD}} + 2 E_{\mathrm{M}}\right)^2+4\left|\tau(\phi)\right|^2]^{3/2}},
\end{equation}
\begin{equation}\label{eq:CQ_eff2}
C_{\mathrm{Q}}^g(\phi)=\frac{2 e^2 \tilde{\alpha}^2\left|g(\phi)\right|^2}{[\left(E_{\mathrm{QD}} - 2 E_{\mathrm{M}}\right)^2+4\left|g(\phi)\right|^2]^{3/2}}.
\end{equation}
Similarly, using Eq.~(\ref{eq:quantum_induc}), we find expressions for the quantum inductance:
\begin{align}
 &  [L_{\mathrm{Q}}^{\tau} (\phi)]^{-1} = \left ( \frac{e}{\hbar}  \right )^2 \nonumber \\
    &\times \frac{2 (|\tau(\phi)|^2)'^2 - [(E_{\mathrm{QD}} + 2 E_{\mathrm{M}})^2 + 4 |\tau(\phi)|^2](|\tau(\phi)|^2)''}{[(E_{\mathrm{QD}} + 2 E_{\mathrm{M}})^2 + 4 |\tau(\phi)|^2]^{3/2}},\label{eq:LQ_eff1} \\
& [L_{\mathrm{Q}}^{g}(\phi)]^{-1} = \left ( \frac{e}{\hbar} \right )^2 \nonumber \\ 
    &\times \frac{2 (|g(\phi)|^2)'^2 - [(E_{\mathrm{QD}} - 2 E_{\mathrm{M}})^2 + 4 |g(\phi)|^2](|g(\phi)|^2)''}{[(E_{\mathrm{QD}} - 2 E_{\mathrm{M}})^2 + 4 |g(\phi)|^2]^{3/2}}, \label{eq:LQ_eff2}
\end{align}
where we introduced the squared moduli
\begin{equation}\label{eq:tau}
|\tau(\phi)|^2 = |\tau_\m{R}|^2 + |\tau_\m{L}|^2 + 2 |\tau_\m{R}| |\tau_\m{L}| \sin (\phi + \phi_\tau),
\end{equation}
\begin{equation}\label{eq:g}
|g( \phi)|^2 = |g_\m{R}|^2 + |g_\m{L}|^2 - 2 |g_\m{R}| |g_\m{L}| \sin (\phi +\phi_g).
\end{equation}
Their dependence on $\phi$ explains the $2 \pi$-periodicity of the quantum capacitance and quantum inductance. 

Our effective-model Hamiltonian conserves fermion parity, and, hence, the $\tau$ and $g$ branches of $C_\m{Q}$ and $L_\m{Q}^{-1}$ correspond to different parity sectors. However, since the effective model neglects bulk states, it does not capture the parity of the full system. The parity of the $\tau$ and $g$ branches can be identified by matching them with the results of the full model. 

The MBSs have symmetric probability density on the left and right ends of the nanowire. Therefore, our Hamiltonian, which has equal absolute values of the tunneling amplitudes on the left and right, implements a well-balanced interferometer. Furthermore, in the case of MBSs with nearly zero energy, $E_\m{M} \approx 0$, the tunneling amplitudes of the normal and Andreev tunneling processes become equal to each other. These two features lead to conditions $|g_\m{R}| = |g_\m{L}| = |\tau_\m{R}| = |\tau_\m{L}|$, and $\phi_g = \phi_\tau$. Substituting these relations into the effective-model Hamiltonian, we restore the standard Majorana operators $\gamma_1 = c + c^\dagger$, $i \gamma_2 = c - c^\dagger$. Furthermore, the difference between $\tau$ and $g$ branches now lies only in the term proportional to $+ \sin (\phi + \phi_\tau)$ in $|\tau(\phi)|^2$, and $- \sin (\phi + \phi_\tau)$ in $|g(\phi)|^2$ , which explains the phase shift $\pi$ between $C_\m{Q}(\phi)$ and $L^{-1}_\m{Q}(\phi)$ in the sectors with even and odd parity.

To compare the results of the effective model, see Eqs.~(\ref{eq:CQ_eff1})–(\ref{eq:LQ_eff2}), with the results obtained numerically, we express the parameters of the simplified Hamiltonian [see Eq.~(\ref{eq:Ham_eff})] in terms of the parameters of the full model [see Eq.~(\ref{eq:Ham_tot})]. The simplified model results from a projection of the full Hamiltonian to the low-energy subspace. Calculating numerically the projections in the second order of perturbation theory in $\lambda_\m{QD}$, we find $E_\m{M} \approx 0$, $E_\m{QD}\approx V_\m{QD} - 0.756\,$meV, $|g_\m{R}| \approx |g_\m{L}| \approx |\tau_\m{R}| \approx |\tau_\m{L}|  \approx 17.5\, \mu$eV, $\phi_\tau = \phi_g = \pi/2$. 
In Fig.~\ref{fig:Majorana_results}(a), we compare the excitation energies of the effective model with the BdG spectrum of the full model.
Figures~\ref{fig:Majorana_results}(c)--\ref{fig:Majorana_results}(f) show that the effective model reproduces $C_\m{Q}(\phi)$ and $L^{-1}_\m{Q}(\phi)$ of the full model.

The results discussed above correspond to the ideal limit of a balanced
interferometer with well-separated zero-energy MBSs, which could be challenging to achieve experimentally. In Appendix~\ref{app:overlap_MBS}, we examine two departures from this limit. First, unequal couplings of the QD to the two nanowire ends make the interferometer unbalanced. In this case, destructive interference is
incomplete, so that the zeros of $C_\m{Q}(\phi)$ are lifted. Second, for overlapping MBSs with finite energy splitting, the two parity branches
generally probe different effective detunings,
$E_\mathrm{QD}\pm2E_\mathrm{M}$, resulting in unequal amplitudes of
$C_\m{Q}$ and $L_\m{Q}^{-1}$ in the even and odd parity sectors, thereby spoiling their exact $\pi$-shift
relation.

Importantly, the same effective-model Hamiltonian $H_\m{eff}$, given in Eq.~(\ref{eq:Ham_eff}), also applies to a non-topological interferometer hosting a single low-energy ABS. Furthermore, as discussed in Sec.~\ref{sec:normal}, under appropriate approximations, the matrices in Eqs.~(\ref{eq:Htau}) and~(\ref{eq:Hg}) together with the resulting expressions for $C_\mathrm{Q}(\phi)$ and $L_\mathrm{Q}^{-1}(\phi)$ also describe the normal interferometer. Thus, the MBS, ABS, and normal scenarios can be mapped onto the same minimal low-energy model. This common effective structure explains the similarity of their flux-dependent responses in certain parameter regimes, as shown below.

\section{Non-topological interferometer}\label{sec:non-topological}

In this section, we focus on the non-topological interferometers that can host either a single ABS or two ABSs. The corresponding configurations are shown in Figs.~\ref{fig:model}(b) and~\ref{fig:model}(c), respectively. Since trivial single-ABS interferometers are known to reproduce Majorana-like quantum-capacitance signatures in certain parameter regimes~\cite{Stanescu2026}, our goal is to examine whether the combined behavior of $C_\mathrm{Q}(\phi)$ and $L_\mathrm{Q}^{-1}(\phi)$ provides a more reliable distinction between topological and trivial interferometers. Additionally, for the two-ABS interferometer, we analyze how the presence of a second low-energy ABS modifies the flux response.

\subsection{Single ABS in the nanowire}\label{sec:1ABS}

\begin{figure}[tb]
    \centering   \includegraphics[width=0.48\textwidth]{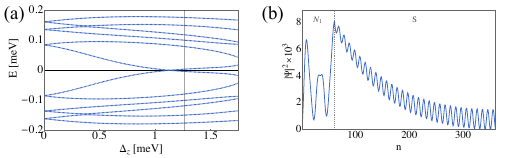}
    \caption{
        (a) Energy spectrum of the nanowire uncoupled from the QD in the non-topological interferometer with one ABS.
        (b) Probability density of the ABS at the Zeeman energy $\Delta_\m{Z} = 1.26\,$meV [indicated by the black line in (a)]. The chosen $\Delta_\m{Z}$, together with a small superconducting gap $\Delta_0 = 0.09\,$meV, results in a finite probability density at the right end of the nanowire, allowing the ABS to couple to the QD from both ends. The parameters are listed in Table~\ref{tab:1ABS} in Appendix~\ref{app:param}.
    }
    \label{fig:1ABS}
\end{figure}

Similarly to the MBS case, we start our investigation by considering the nanowire decoupled from the QD. In Fig.~\ref{fig:1ABS}(a), the energy spectrum of the nanowire shows the presence of a superconducting gap and a single in-gap ABS. We choose a relatively small superconducting gap, $\Delta_0 = 0.09\,$meV, which leads to a finite probability density, $|\Psi|^2$, at the right end of the nanowire; see Fig.~\ref{fig:1ABS}(b).
As a result, the ABS can couple to the QD from both the left and right sides \cite{Hess2021}.
A similar effect can be achieved by considering a larger value of $\Delta_0$ together with a smaller $N_s$ (short superconducting section).  

\begin{figure}[tb]
    \centering   \includegraphics[width=0.48\textwidth]{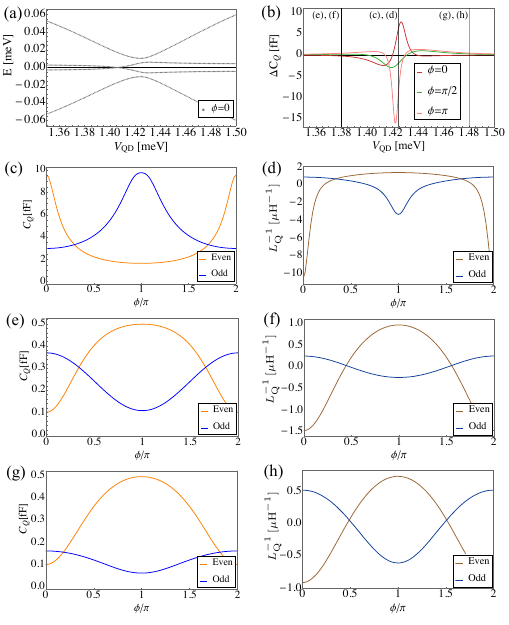}
    \caption{(a) Low-energy spectrum of the non-topological interferometer with one ABS. (b) Difference between quantum capacitance in the even and odd parity sectors $\Delta C_\m{Q}$ as a function of $V_\m{QD}$. (c), (e), (g) Even and odd branches of the quantum capacitance $C_\m{Q}$ and (d), (f), (h) quantum inductance $L^{-1}_\m{Q}$ at $V_\m{QD} = 1.425\,$meV, $V_\m{QD} = 1.38\,$meV and $V_\m{QD} = 1.48\,$meV [indicated by the black lines in (b)].
    In this parameter regime, $C_\m{Q}(\phi)$ and $L_\m{Q}^{-1}(\phi)$ show Majorana-like behavior, namely $2\pi$ periodicity and nearly $\pi$ phase shift.
    The parameters are listed in Table~\ref{tab:1ABS} in Appendix~\ref{app:param}.}
    \label{fig:1ABS_results}
\end{figure}

In the next step, we couple the nanowire to the QD, focusing on the weak-coupling regime. As discussed in Sec.~\ref{sec:topological}, MBSs exhibit a symmetric probability density, resulting in equal absolute values of effective tunneling amplitudes at the left and right ends of the nanowire. In contrast, the ABS is spatially asymmetric and more localized near the left end, leading to unequal effective couplings to the QD. To make the interferometer well-balanced (similar to the MBS case), we introduce asymmetric tunneling amplitudes, choosing $\lambda_\m{L} = 0.01$ and $\lambda_\m{R} = 0.03$. For this setup, in Fig.~\ref{fig:1ABS_results}(b), we plot the difference between the quantum capacitance in the even and odd parity sectors $\Delta C_Q$ for the values of $V_\m{QD}$ near the avoided crossing between the ABS and QD energy levels [see Fig.~\ref{fig:1ABS_results}(a)].

Next, we plot the quantum
capacitance and quantum inductance in the even and odd
parity sectors as functions of $\phi$ for three representative values of $V_\m{QD}$, and compare the ABS results with the MBS ones. For the first value near the resonances, $V_\m{QD} = 1.425\,$meV, the results are presented in Figs.~\ref{fig:1ABS_results}(c) and~\ref{fig:1ABS_results}(d). We see that both $C_\m{Q}(\phi)$ and $L^{-1}_\m{Q}(\phi)$ exhibit a $2 \pi$ periodicity, however, only $C_\m{Q}(\phi)$ shows an almost exact $\pi$ shift between the even and odd branches. The shape of the $C_\mathrm{Q}(\phi)$ curves closely resembles that of the unbalanced topological interferometer shown in Appendix~\ref{app:overlap_MBS}. A similar situation is presented in Figs.~\ref{fig:1ABS_results}(e) and~\ref{fig:1ABS_results}(f) for $V_\m{QD} = 1.38\,$meV, where $C_\m{Q}(\phi)$ is almost identical to the MBS result shown in Fig.~\ref{fig:Majorana_results}(e), while $L^{-1}_\m{Q}(\phi)$ is slightly different. The adjustment of the value of $V_\m{QD}$ to $1.48\,$meV makes the behavior of $L^{-1}_\m{Q}(\phi)$ very similar to the corresponding MBS result. 

In summary, in some parameter regimes, the ABS response in the trivial phase is qualitatively similar to the MBS response in the topological phase. In both Figs.~\ref{fig:Majorana_results} and~\ref{fig:1ABS_results}, the quantum capacitance and quantum inductance show a $2\pi$ periodicity. The even and odd ABS branches are not perfectly shifted by $\pi$, but the deviation is sufficiently small that, within the current experimental resolution~\cite{1Microsoft}, they may be difficult to distinguish from the MBSs in an experiment, even if one measures both $C_\m{Q}(\phi)$ and $L^{-1}_\m{Q}(\phi)$. Generally speaking, this behavior is not universal across all parameter values. In Appendix~\ref{app:not_mimic}, we provide examples where the ABS exhibits Majorana-like behavior only in $C_\mathrm{Q}(\phi)$, or only in $L_\mathrm{Q}^{-1}(\phi)$, or neither of them.

In the weak-coupling regime, our numerical results can be understood using the simplified model given by Eq.~(\ref{eq:Ham_eff}). The ABS quantum capacitance and quantum inductance are governed by the same expressions as in the MBS case, Eqs.~(\ref{eq:CQ_eff1})-(\ref{eq:LQ_eff2}), which explains the emergence of Majorana-like behavior in the trivial phase for certain parameter regimes. In the ABS case, however, we consider a relatively small superconducting gap, which complicates the extraction of the effective parameters of the simplified model. Nevertheless, we checked numerically that for the higher value of $\Delta_0$, the effective model (with parameters obtained within second-order perturbation theory) agrees well with the full model; see Appendix~\ref{app:not_mimic}.

\subsection{Two ABSs in the nanowire}

\begin{figure}[b]
    \centering   \includegraphics[width=0.48\textwidth]{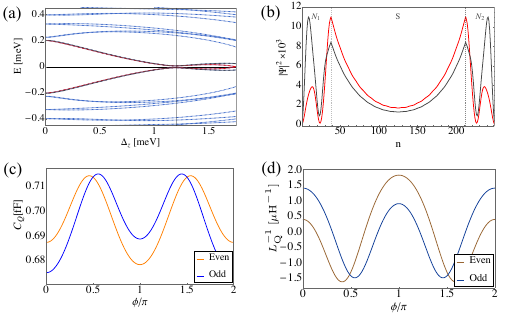}
    \caption{(a) Energy spectrum of the nanowire uncoupled from the QD in the non-topological interferometer with two ABSs, where the bulk states are shown in blue, one ABS in red, and another in gray. (b) Probability density of two ABSs at the Zeeman energy $\Delta_\m{Z} = 1.2\,$meV [indicated by the black line in (a)]. (c) Even and odd branches of the quantum capacitance $C_\m{Q}$ and (d) quantum inductance $L^{-1}_\m{Q}$ at $V_\m{QD} = 1.261\,$meV. In contrast to MBSs, two delocalized ABSs with nearly zero energy can transfer a Cooper pair of charge $2e$ that results in substantial $\pi$ components in $C_\m{Q}(\phi)$ and $
    L_\m{Q}^{-1}(\phi)$. This drastically distinguishes this case from the MBS scenario. The parameters are listed in Table~\ref {tab:2ABS} in Appendix~\ref{app:param}.}
    \label{fig:2ABS_pinned}
\end{figure}

We now examine the setup shown in Fig.~\ref{fig:model}(c), where two normal sections are attached to the superconducting nanowire:
    one on the left and another on the right side.
As a result, the nanowire hosts two ABSs localized near the boundaries between the superconducting and normal sections.
We focus on the regime where both ABS energies are close to zero in the absence of the QD [see Fig.~\ref{fig:2ABS_pinned}(a)].
Furthermore, we choose identical parameters for the left and right normal sections, resulting in a symmetric probability density of the ABSs, as shown in Fig.~\ref{fig:2ABS_pinned}(b). In this respect, the situation resembles the MBS case, where the probability density is also symmetric.
However, in contrast to MBSs, the ABSs considered here have a significant overlap, enabling them to couple to the QD from both ends.
Another important difference is that two ABSs can transfer charge of $2e$, i.e., a Cooper pair, that results in substantial $h/(2 e)$ components in flux or equivalently $\pi$ components in phase $\phi$.
Consequently, both even and odd branches of $C_\mathrm{Q}(\phi)$ and $L_\mathrm{Q}^{-1}(\phi)$ exhibit almost $\pi$-periodicity [see Figs.~\ref{fig:2ABS_pinned}(c) and~\ref{fig:2ABS_pinned}(d)], which distinguishes this case from the MBS scenario. However, the energy of one of the ABSs can be tuned closer to the bulk states.
This ABS effectively mimics the edge of the bulk gap, and the system crosses over to behavior similar to the single-ABS case. Therefore, this interferometer, similar
to the interferometer with only one ABS, can also exhibit
Majorana-like behavior; see Appendix~\ref{app:2ABS}.

The results of the two-ABS case can also be understood within a simplified model. This requires extending Eq.~(\ref{eq:Ham_eff}) to include a second ABS level and its corresponding coupling to the QD. However, in this extended model, the analytical expressions for $C_\mathrm{Q}(\phi)$ and $L_\mathrm{Q}^{-1}(\phi)$ become rather cumbersome, making it difficult to extract a clear physical intuition for their dependence on the system parameters.

\section{Normal interferometer}\label{sec:normal}

Finally, we discuss the normal interferometer in the absence of a superconductor that is shown in Fig.~\ref{fig:model}(d). This case is relevant for two reasons. First, it is of fundamental interest because it provides an understanding of the behavior of the quantum capacitance and quantum inductance in a purely normal system. Second, this case is of practical importance. In experiments, superconductivity in a nanowire is usually induced via the proximity effect, which may not always behave as expected or may be weakened under certain conditions. In such situations, the nanowire can remain normal, with neither MBSs nor ABSs, or with a negligibly small superconducting gap. This can lead to a potential misinterpretation of the observed signals. 

For simplicity, in this section, we restrict the model to the minimal ingredients needed to capture the essential physics, that is, we set the superconducting gap and spin–orbit coupling to zero while keeping a finite Zeeman energy to lift the spin degeneracy.
Nevertheless, the same behavior persists in the presence of a sufficiently small superconducting gap and/or finite spin–orbit coupling, provided that the Zeeman energy dominates over the relevant energy scales.

\begin{figure}[tb]
    \centering   \includegraphics[width=0.48\textwidth]{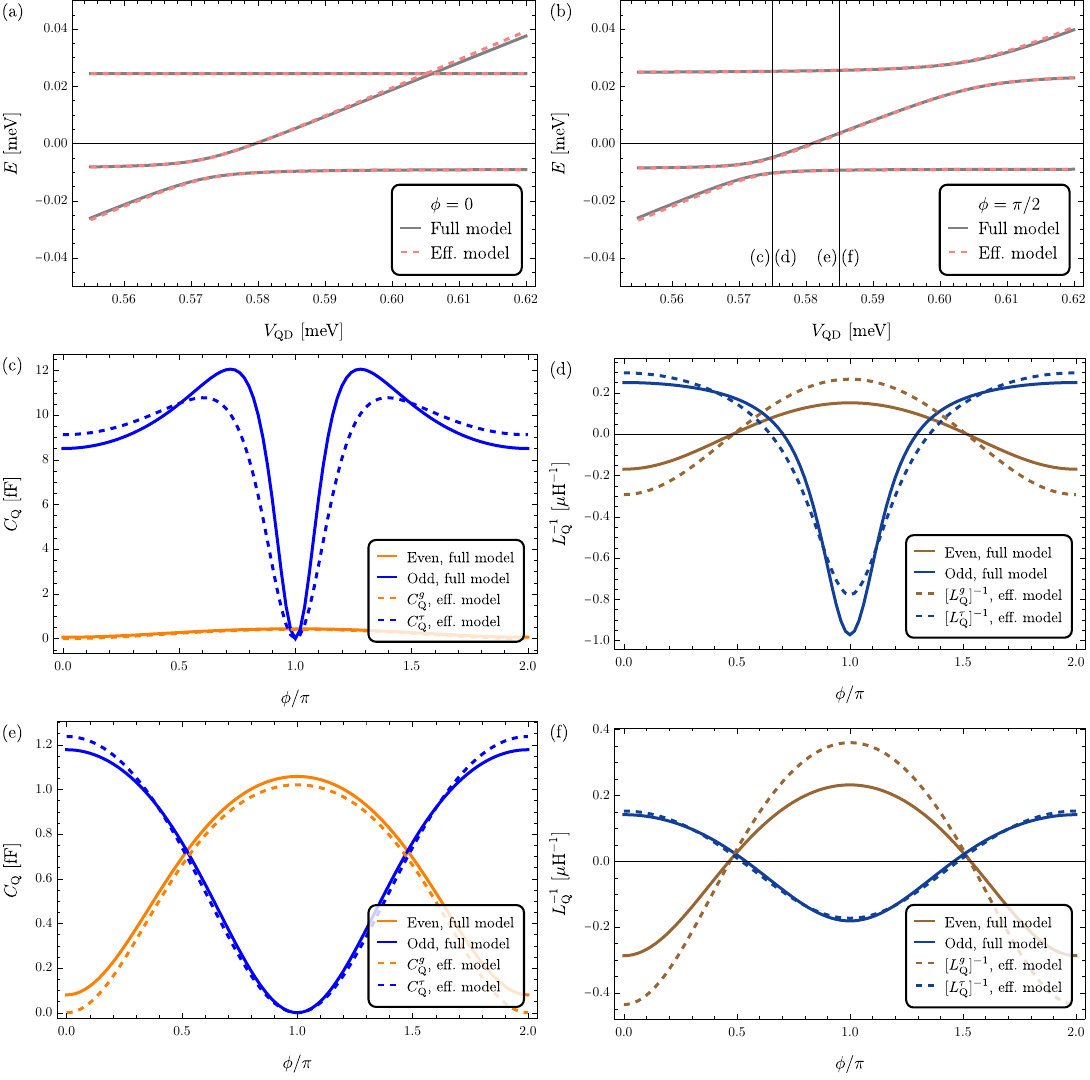}
    \caption{(a), (b) Low-energy single-particle spectrum of the normal interferometer in dependence on $V_\m{QD}$ for $\phi = 0$, and $\phi = \pi/2$. (c), (e) Even and odd branches of the quantum capacitance $C_\m{Q}(\phi)$ and (d), (f) quantum inductance $L^{-1}_\m{Q}(\phi)$ at $V_\m{QD} = 0.575\,$meV and $V_\m{QD} = 0.585\,$meV [indicated by the black lines in (b)]. For $V_\mathrm{QD}=0.575\,\mathrm{meV}$, the response is dominated by the $\tau$ branch, while the $g$ branch is suppressed by the large detuning. For $V_\mathrm{QD}=0.585\,\mathrm{meV}$, both branches have comparable amplitudes and exhibit a Majorana-like $2\pi$ periodicity with an approximate $\pi$ phase shift, despite the system being purely normal. The effective model (dashed lines) is in good agreement with the full Hamiltonian (solid lines). The parameters are listed in Table~\ref{tab:normal} in Appendix~\ref{app:param}.}
    \label{fig:normal_results}
\end{figure}

In the absence of the QD, the energy levels of the normal nanowire are given by $E_j^\pm = -\mu + 2 t \,\{1 - \cos[ \pi j/(N+1)]\} \pm \Delta_\m{Z}$, where $j =1,2,\dots$ labels the energy levels and $\pm$ corresponds to different spins. 
Note that since the superconducting gap is zero, we present here the non-BdG single-particle energies.
The probability density of any of these uncoupled nanowire states is delocalized along the entire nanowire, where the position wave function is given by $\Psi_{j,n} = \sqrt{2/(N+1)} \sin[\pi j n/ (N+1)]$ with the site index $n=1,2,\dots,N$.
As a result, once the coupling to the QD is switched on, the nanowire states can couple to the QD from both the left and right ends.
Specifically, to first order, the effective coupling of the QD to the left side of a nanowire eigenstate is $-e^{i \phi} \lambda_\m{QD} t \Psi_{j,1}$, and to the right side $-\lambda_\m{QD} t \Psi_{j,N}$.
Since $\m{sign}(\Psi_{j,N})$ changes for consecutive $j$, the interference of the left and right sides can be exactly constructive or destructive in the special case $\phi \in \{0, \pi\}$
    [see Fig.~\ref{fig:normal_results}(a)].
Note that the amplitude of the couplings changes for different $j$
    [see Fig.~\ref{fig:normal_results}(b)].

For a better understanding of the numerical results shown in Figs.~\ref{fig:normal_results}(c)--\ref{fig:normal_results}(f), we first introduce a low-energy effective model for the normal interferometer.
We assume that the high-energy states do not contribute to the phase dependence and only renormalize the potential of the QD.
To obtain a nonzero response in both parity sectors, the minimal low-energy description now requires two nanowire energy levels, denoted by $E_\tau$ and $E_g$. These levels are coupled to the QD level $E_\mathrm{QD}$ through normal tunneling processes with amplitudes $\tau(\phi)$ and $g(\phi)$, respectively. The tunneling amplitudes are described by the same expressions as in the MBS case, with squared moduli given by Eqs.~(\ref{eq:tau}) and~(\ref{eq:g}).
However, in the present case, both amplitudes correspond to normal tunneling processes. The resulting effective-model Hamiltonian takes the form
\begin{equation}
    H = E_\m{QD} d^{\dagger} d  + \sum_{j \in \{\tau,g\}} E_j c^{\dagger}_j c_j  + [j(\phi) d^{\dagger} c_j + \text{H.c.}],
\end{equation}
where $c_j^\dagger$ and $c_j$ [$d^\dagger$ and $d$] are fermionic creation and annihilation operators for the nanowire level $j\in \{\tau,g \}$ [the QD]. We tune the QD level so that it lies between the two nanowire levels $E_\tau < E_\mathrm{QD} < E_g$, allowing interaction of the QD with both $E_\tau$ and $E_g$.

The Hamiltonian $H$ is block diagonal and consists of four independent particle-number sectors in the Fock basis $|x y z\rangle = |x \rangle_\tau |y \rangle_\m{QD} |z \rangle_g$ with $x, y, z \in \{0,1\}$. The sectors with zero particles, $|000\rangle$, and three particles, $|111\rangle$, are trivial, since both the quantum capacitance and quantum inductance vanish in these cases. The nontrivial contributions arise from the one- and two-particle sectors. The one-particle sector is spanned by the basis $\{|100\rangle, |010\rangle, |001\rangle \}$ with the corresponding Hamiltonian
\begin{equation}\label{eq:H1}
    H_1 =
    \left(
    \begin{array}{c|c}
        \begin{matrix}
            E_\tau & \tau^*(\phi) \\
            \tau(\phi) & E_\mathrm{QD}
        \end{matrix}
        &
        \begin{matrix}
            0 \\
            g(\phi)
        \end{matrix}
        \\ \hline
        \begin{matrix}
            \quad 0 & \quad \: \:  g^*(\phi)
        \end{matrix}
        &
        E_g
    \end{array}
    \right).
\end{equation}
The two-particle Hamiltonian $H_2$ is written in the basis $ \{|110\rangle, |101\rangle, |011\rangle \}$ as
\begin{equation}\label{eq:H2}
H_2 =
\left(
\begin{array}{c|c}
\begin{matrix}
E_\tau + E_\m{QD} &  g(\phi) \\
g^*(\phi) & E_\tau + E_g 
\end{matrix}
&
\begin{matrix}
0 \\
\tau^*(\phi)
\end{matrix}
\\ \hline
\begin{matrix}
\quad 0 \quad \,  &  \quad \quad \quad \tau(\phi)
\end{matrix}
&
E_\m{QD} + E_g

\end{array}
\right).
\end{equation}
The last basis state in each matrix affects the ground-state energy only through virtual transitions to the low-energy sector and therefore contributes only in second order in the corresponding coupling.
For the parameters considered here, this correction is much smaller than the characteristic low-energy scales.
Therefore, the ground-state energy is well captured by the reduced $2\times2$ blocks, which acquire the same structure as in the MBS case [see Eqs.~(\ref{eq:Htau}) and~(\ref{eq:Hg})]. Finding the lowest energies of each block, we calculate the quantum capacitance and quantum inductance:
\begin{align}
&C_{\mathrm{Q}}^{j}(\phi)=\frac{2 e^2 \tilde{\alpha}^2\left|j(\phi)\right|^2}{[\left(E_{\mathrm{QD}} - E_j\right)^2+4\left|j(\phi)\right|^2]^{3/2}},\\
&  [L_{\mathrm{Q}}^{j} (\phi)]^{-1} = \left ( \frac{e}{\hbar}  \right )^2 \nonumber \\
    &\times \frac{2 (|j(\phi)|^2)'^2 - [(E_{\mathrm{QD}} - E_j)^2 + 4 |j(\phi)|^2](|j(\phi)|^2)''}{[(E_{\mathrm{QD}} -E_j)^2 + 4 |j(\phi)|^2]^{3/2}},
\end{align}
where $j = \tau$ corresponds to $H_1$ matrix, and $j = g$ to $H_2$ matrix. Thus, the quantum-capacitance and quantum-inductance expressions take the same form as in the MBS case [Eqs.~(\ref{eq:CQ_eff1})–(\ref{eq:LQ_eff2})], with $(E_\m{QD}+2E_\m{M})^2$ replaced by $(E_\m{QD}-E_\tau)^2$ in the $\tau$ branch and $(E_\m{QD}-2E_\m{M})^2$ replaced by $(E_g-E_\m{QD})^2$ in the $g$ branch.

In Fig.~\ref{fig:normal_results}, we compare this effective model (dashed lines) with the full model (solid lines) by
projecting the full Hamiltonian onto the low-energy subspace within
second-order perturbation theory.
This yields $E_\m{QD}\approx V_\m{QD} - 0.58\,$meV, $E_\tau \approx -8.9\, \mu$eV, $E_g \approx24.4\,\mu$eV, $|\tau_\m{R}| = |\tau_\m{L}| \approx 1.7\, \mu$eV, $|g_\m{R}| =|g_\m{L}| \approx 3.5\, \mu$eV, $\phi_\tau = \pi/2$, and $\phi_g =\pi/2 $. 

To analyze the dependence of the quantum capacitance and quantum inductance on $\phi$, we first choose $V_\mathrm{QD}=0.575\,\mathrm{meV}$, corresponding to
$E_\mathrm{QD}\approx -5\,\mu\mathrm{eV}$, so that the QD level lies
close to the lower nanowire level $E_\tau$. The results are shown in
Figs.~\ref{fig:normal_results}(c) and~\ref{fig:normal_results}(d). In this regime, the $g$-branch response is suppressed relative to the $\tau$ branch
because the denominators of $C_\m{Q}^g(\phi)$ and of $[L_\m{Q}^g(\phi)]^{-1}$ contain the large detuning $E_g-E_\mathrm{QD}$. This asymmetry between the branch amplitudes is similar to that produced by a finite energy splitting between overlapping MBSs; see Appendix~\ref{app:overlap_MBS}.

To obtain comparable amplitudes in the two parity sectors, we tune
the QD level closer to the middle of the interval between $E_\tau$ and
$E_g$ by choosing $V_\mathrm{QD}=0.585\,\mathrm{meV}$ [black line in
Fig.~\ref{fig:normal_results}(b)]. In this case, both detunings
$E_\mathrm{QD}-E_\tau$ and $E_g-E_\mathrm{QD}$ are of comparable size.
As shown in Figs.~\ref{fig:normal_results}(e) and
\ref{fig:normal_results}(f), the even and odd branches then acquire similar amplitudes and display Majorana-like behavior. We thus can conclude that the $2\pi$ periodicity and $\pi$ phase shift in the quantum capacitance and quantum inductance are not sufficient conditions to identify the interferometer as topological, or even as superconducting. 

Note that in the normal interferometer, tuning $V_\m{QD}$ is analogous to choosing the QD operating point in capacitance-based readout setups.
However, obtaining a Majorana-like response requires an additional condition: the chemical potential must place the midpoint between the two relevant normal levels sufficiently close to zero energy.

Finally, we numerically verified that the qualitative behavior of the quantum capacitance and quantum inductance is robust against onsite potential disorder, provided that $\mu$ and $V_\m{QD}$ are adjusted accordingly; see Appendix~\ref{app:disorder}.

\section{Conclusions and outlook}\label{sec:conclusion}

We considered interferometric nanowire–quantum-dot setups with four different types of nanowires: a topological one hosting MBSs, the superconducting nanowires in the non-topological phase with one or two ABSs, and a fully normal nanowire. For all configurations, we calculated the parity-resolved quantum capacitance $C_\m{Q}$ and quantum inductance $L^{-1}_\m{Q}$ in dependence on the magnetic flux, or equivalently the phase difference $\phi$, by exact diagonalization of the tight-binding model.

Our results demonstrate that signatures associated with zero-energy MBSs, such as $2\pi$ periodicity and a relative $\pi$ phase shift between even and odd parity sectors in $C_\mathrm{Q}(\phi)$ and $L_\mathrm{Q}^{-1}(\phi)$, are not unique to topological systems. Within the current experimental resolution~\cite{1Microsoft}, similar Majorana-like behavior can arise from trivial ABSs and even from purely normal nanowires in certain realistic parameter regimes. This highlights that capacitance and inductance-based interferometry, while powerful, does not provide a definitive probe of topological superconductivity.

To gain physical insight, we introduced a simplified effective model obtained by projecting the full Hamiltonian onto the low-energy subspace, which reproduces the numerical BdG results with good accuracy. We found that under appropriate approximations, all cases considered in this work (except for two nearly zero-energy ABSs) can be mapped onto the same minimal low-energy model. This common effective structure explains why the MBS, single-ABS, and normal scenarios can exhibit similar flux-dependent responses despite their different physical origins. 

We also showed that the Majorana-like behavior in non-topological interferometers is not universal and depends sensitively on the system parameters. In certain parameter regimes, it is therefore possible to determine that the interferometer is not in the topological phase. For example, in the topologically trivial phase with two nearly zero-energy ABSs, which can transfer a Cooper pair, the response exhibits a pronounced $\pi$-periodicity, in contrast to the $2\pi$-periodicity of the MBSs in the topological phase.

We had two motivations for this study. First, it is of fundamental interest, as it provides insight into the behavior of quantum capacitance and quantum inductance in systems with different underlying physical properties.
Second, our results have practical implications for the interpretation of capacitance- and inductance-based parity-readout experiments.
In particular, it is important to ensure that a sufficiently large superconducting gap is induced in the nanowire via the proximity effect. If the induced gap is too small or not fully developed, the system may remain effectively normal, potentially leading to misinterpretation of the observed signals as signatures of topological physics. Furthermore, even in the presence of superconductivity, the observed response may originate from trivial ABSs rather than from MBSs, making the identification of topological signatures ambiguous.

\acknowledgments
We thank Sebastian Miles for fruitful discussions on the construction of the effective model. This work was supported as part of NCCR SPIN, a National Center of Competence in Research, funded by
the Swiss National Science Foundation (grant number 225153). 
DL acknowledges the Deanship of Research at KFUPM and the Quantum Center for the support received under Grant no. CUP25102 and INQC2600.

\appendix

\section{Nonideal topological interferometers}\label{app:overlap_MBS}

\begin{figure}[tb]
    \centering   \includegraphics[width=0.48\textwidth]{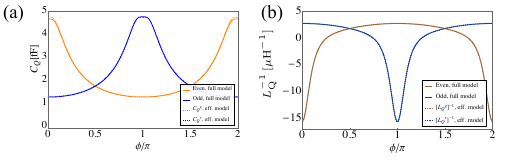}
    \caption{Even and odd branches of (a) the quantum capacitance $C_\m{Q}$ and (b) the quantum inductance $L_\m{Q}^{-1}$ in an unbalanced topological interferometer. The nanowire hosts the same well-separated, nearly zero-energy MBSs as in Fig.~\ref{fig:Majorana}, but the QD couplings to the two nanowire ends are unequal, $\lambda_\m{L}=0.015$ and $\lambda_\m{R}=0.02$, which lifts the zeros of $C_\m{Q}(\phi)$. The effective-model results (dashed lines) agree with the full-model results (solid lines). The parameters are listed in Table~\ref{tab:Majorana} in Appendix~\ref{app:param}. 
    }
    \label{fig:MBS_unbal}
\end{figure}

In the main text, we focused on the ideal limit of a balanced interferometer hosting well-separated MBSs with negligible energy splitting. Here, we relax these two assumptions separately.

First, we use the same long topological nanowire as in Fig.~\ref{fig:Majorana}, for which the
MBSs are well separated and $E_\m{M}\approx0$, but choose unequal QD--nanowire couplings, $\lambda_\m{L}=0.015$ and $\lambda_\m{R}=0.02$. The resulting interferometer is therefore unbalanced. As shown in Fig.~\ref{fig:MBS_unbal}, both $C_\m{Q}(\phi)$ and $L_\m{Q}^{-1}(\phi)$ remain $2\pi$-periodic, and the two parity branches remain related by a $\pi$ shift. However, the unequal couplings make destructive interference incomplete, lifting the zeros of \(C_\m{Q}(\phi)\) present in the balanced case. The effective model reproduces the full-model results with \(E_\m{M}\approx0\), \(E_\m{QD}\approx V_\m{QD}-0.739~\mathrm{meV}\), \(|g_\m{R}|\approx|\tau_\m{R}|\approx17.6~\mu\mathrm{eV}\), \(|g_\m{L}|\approx|\tau_\m{L}|\approx13.2~\mu\mathrm{eV}\), and \(\phi_\tau=\phi_g=\pi/2\).

\begin{figure}[tb]
    \centering   \includegraphics[width=0.48\textwidth]{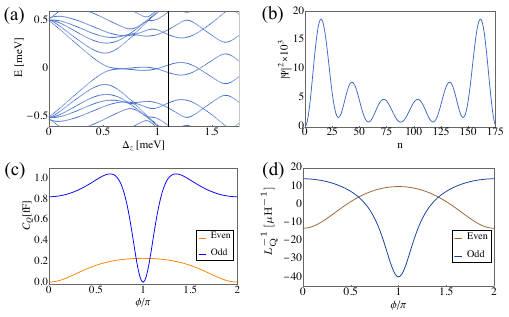}
    \caption{(a) Energy spectrum of the shorter topological nanowire decoupled from the QD. (b) Probability density of the lowest-energy state at $\Delta_\m{Z} = 1.1\,$meV [indicated by the black line in (a)]. (c) Even and odd branches of the quantum capacitance $C_\m{Q}$ and (d) the quantum inductance $L_\m{Q}^{-1}$. Both responses remain $2\pi$-periodic but the two parity branches have different amplitudes and are no longer related by $\pi$ shift because of the finite-energy MBS splitting. The parameters are listed in Table~\ref{tab:Majorana} in Appendix~\ref{app:param}.   
}
    \label{fig:MBS_overlap}
\end{figure}

Second, we consider a shorter topological nanowire, where the MBSs have a finite-energy splitting and a finite overlap. Figures~\ref{fig:MBS_overlap}(a) and \ref{fig:MBS_overlap}(b) show, respectively, the spectrum of the nanowire decoupled from the QD and the MBS probability density. After coupling the nanowire to the QD, we calculate \(C_\m{Q}(\phi)\) and \(L_\m{Q}^{-1}(\phi)\) at \(V_\m{QD}=1.25~\mathrm{meV}\). As shown in Figs.~\ref{fig:MBS_overlap}(c) and \ref{fig:MBS_overlap}(d), both responses remain \(2\pi\)-periodic, but the parity branches acquire substantially different amplitudes and are no longer related by a perfect \(\pi\) shift. Within the effective model of Sec.~\ref{sec:topological}, the two branches depend on the detunings \(E_\m{QD}+2E_\m{M}\) and \(E_\m{QD}-2E_\m{M}\) [see Eqs.~(\ref{eq:CQ_eff1})--(\ref{eq:LQ_eff2})]. Away from \(E_\m{QD}=0\), these detunings generally have different magnitudes, so one branch can lie closer to resonance and produce a larger response.

These results show that deviations from the ideal response of Fig.~\ref{fig:Majorana_results} do not
by themselves indicate a topologically trivial system. An unbalanced topological interferometer can exhibit lifted capacitance minima, similar to ABS case in Fig.~\ref{fig:1ABS}, while overlapping MBSs with finite energy splitting can produce strongly unequal parity-branch amplitudes, similar to the normal-interferometer response in Fig.~\ref{fig:normal_results}.

\section{Example of the ABS response. Comparison of the full and simplified models}\label{app:not_mimic}

In this Appendix, we examine the single-ABS case in a parameter regime where the behavior of the quantum capacitance and quantum inductance differs from the MBS scenario.
In contrast to the example discussed in the main text [Fig.~\ref{fig:1ABS_results}], we now generate a delocalized ABS by considering a larger superconducting gap, $\Delta_0 = 0.25\,$meV, together with a shorter superconducting section, $N_s = 175$ (all parameters are listed in Table~\ref{tab:1ABS} in Appendix~\ref{app:param}). We also consider  $\lambda_\mathrm{QD} = 0.02$ on both the left and right sides.

\begin{figure}[tb]
    \centering   \includegraphics[width=0.48\textwidth]{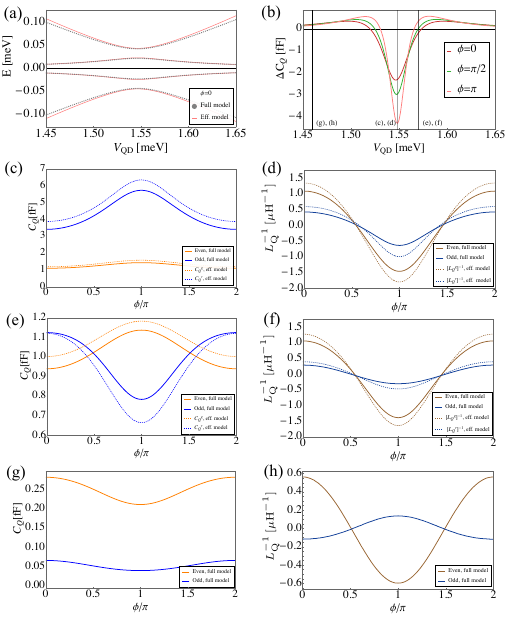}
    \caption{(a) Low-energy spectrum of the non-topological interferometer with one ABS. (b) Difference between quantum capacitance in the even and odd parity sectors, $\Delta C_\m{Q}$, as a function of $V_\m{QD}$. (c), (e), (g) Even and odd branches of the quantum capacitance $C_\m{Q}(\phi)$ and (d), (f), (h) quantum inductance $L^{-1}_\m{Q}(\phi)$. (c), (d) At the resonance value of $V_\m{QD} = 1.548\,$meV [indicated by the black line in (b)], neither $C_\m{Q}$ nor $L^{-1}_\m{Q}$ shows the Majorana-like behavior. (e), (f) Tuning $V_\m{QD}$ to the value $1.57\,$meV makes $C_\m{Q}$ very similar to MBS case, while $L^{-1}_\m{Q}$ still differs, (g),(h) while at $V_\m{QD} = 1.46\,$meV, it is vice versa. The effective model [dashed lines in (c)-(f)] is in good agreement with the full Hamiltonian (solid lines). 
    The parameters are listed in Table~\ref{tab:1ABS} in Appendix~\ref{app:param}.}
    \label{fig:1ABS_not_mimic}
\end{figure}

The results are shown in Fig.~\ref{fig:1ABS_not_mimic}. Depending on the chosen value of $V_\mathrm{QD}$ [indicated by black lines in Fig.~\ref{fig:1ABS_not_mimic}(b)], the ABS exhibits Majorana-like behavior either only in $C_\mathrm{Q}(\phi)$, only in $L_\mathrm{Q}^{-1}(\phi)$, or in neither quantity. In this regime, measuring a single observable may be insufficient to distinguish between trivial and topological behavior.
However, simultaneous measurements of both $C_\mathrm{Q}(\phi)$ and $L_\mathrm{Q}^{-1}(\phi)$ allow one to identify the interferometer as trivial.
Nevertheless, for parameter regimes such as those shown in Fig.~\ref{fig:1ABS_results}, even measuring both $C_\mathrm{Q}(\phi)$ and $L_\mathrm{Q}^{-1}(\phi)$ may not be sufficient to distinguish between the topological and trivial phases.

We then compare the simplified effective model, defined in Eq.~(\ref{eq:Ham_eff}), with the full model, given by Eq.~(\ref{eq:Ham_tot}). We project the full Hamiltonian onto the low-energy subspace in the second order of perturbation theory. From this projection, we extract the effective parameters of the simplified model: $E_\m{M} \approx 0$, $E_\m{QD} \approx V_\m{QD} - 1.55 \,$meV, $|g_\m{R}| \approx 4.2\,\mu$eV, $|g_\m{L}| \approx 29.5\,\mu$eV, $|\tau_\m{R}| \approx 2\,\mu$eV, $|\tau_\m{L}| \approx 8.2\,\mu$eV, $\phi_\tau = \pi/2$, and $\phi_g = 3 \pi/2$. As shown in Figs.~\ref{fig:1ABS_not_mimic}(a), and~\ref{fig:1ABS_not_mimic}(c)--\ref{fig:1ABS_not_mimic}(f), the effective model agrees well with the full model.
However, the deviations are slightly larger than in the MBS case [Fig.~\ref{fig:Majorana}].
We attribute this to the smaller superconducting gap, which is reduced by a factor of two compared to the MBS case.
As a result, the second-order perturbation theory used to derive the effective-model parameters is no longer accurate enough. In the subplots (g) and (h), the QD-level is close to the continuum, so the effective model is not applicable in this case; for this reason, we show only the full-model results there.

From the extracted effective-model parameters, we can identify several important differences compared to the MBS case. First, the normal tunneling amplitudes $|g_i|$ and the Andreev amplitudes $|\tau_i|$ are unequal. Second, the ABS is more localized near the left end of the nanowire, which leads to asymmetric effective couplings, i.e., $|g_\mathrm{L}| > |g_\mathrm{R}|$ and $|\tau_\mathrm{L}| > |\tau_\mathrm{R}|$. Finally, in Eqs.~(\ref{eq:tau})~and~(\ref{eq:g}) for the squared moduli, we have the term $\sim+\sin(\phi + \phi_\tau)$ in $|\tau(\phi)|^2$ , and the term $\sim-\sin(\phi + \phi_g)$ in $|g(\phi)|^2$. Taking into account the phase offsets $\phi_\tau = \pi/2$ and $\phi_g = 3\pi/2$, one finds that these contributions are in phase. This explains why, for most parameter regimes, no $\pi$ phase shift appears and the even and odd branches remain aligned, however, the dependence on the phase $\phi$ is generally more involved.

\section{Non-topological interferometer with two ABSs mimics the single ABS interferometer}\label{app:2ABS}

\begin{figure}[tb]
    \centering   \includegraphics[width=0.48\textwidth]{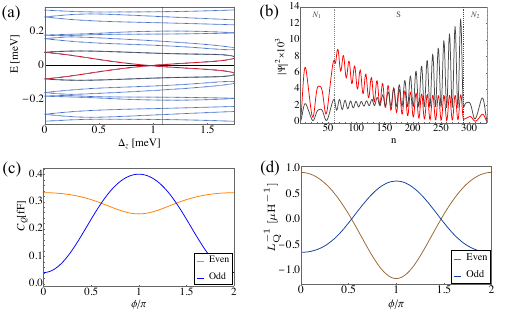}
    \caption{(a) Energy spectrum of the nanowire uncoupled from the QD in the non-topological interferometer with two ABSs, where the bulk states are shown in blue, one ABS in red, and another in gray. (b) Probability density of two ABSs at the Zeeman energy $\Delta_\m{Z} = 1.08\,$meV [indicated by the black line in (a)]. (c) Even and odd branches of the quantum capacitance $C_\m{Q}(\phi)$ and (d) quantum inductance $L^{-1}_\m{Q}(\phi)$ at $V_\m{QD} = 1.1\,$meV. The energy of the second ABS (gray) is tuned closer to the bulk, mimicking the edge of the bulk gap. As a result, the system behaves similarly to the single ABS case shown in Fig.~\ref{fig:1ABS_results}. The parameters are listed in Table~\ref {tab:2ABS} in Appendix~\ref{app:param}.}
    \label{fig:2ABS_bulky}
\end{figure}

In this section, we focus on the non-topological interferometer that hosts two ABSs.
We choose the parameters of the nanowire decoupled from the QD such that the energy of one of the ABSs is tuned closer to the edge of the bulk gap, while the energy of the second ABS stays close to zero; see Fig.~\ref{fig:2ABS_bulky}(a).
As a result, the higher-energy ABS [shown in gray in Figs.~\ref{fig:2ABS_bulky}(a) and~\ref{fig:2ABS_bulky}(b)] effectively mimics a bulk state, and the system can be viewed as equivalent to the single-ABS setup discussed in Sec.~\ref{sec:1ABS}, with only one low-energy ABS.
Figures~\ref{fig:2ABS_bulky}(c) and~\ref{fig:2ABS_bulky}(d) show the corresponding quantum capacitance and quantum inductance in dependence on the phase $\phi$, which are in good agreement with the single-ABS results presented in Fig.~\ref{fig:1ABS_results} of the main text.
Thus, this interferometer can also exhibit Majorana-like behavior, similar to the single-ABS case.

\section{Normal interferometer with disorder}\label{app:disorder}

\begin{figure*}[]
    \centering
    \includegraphics[width=\linewidth]{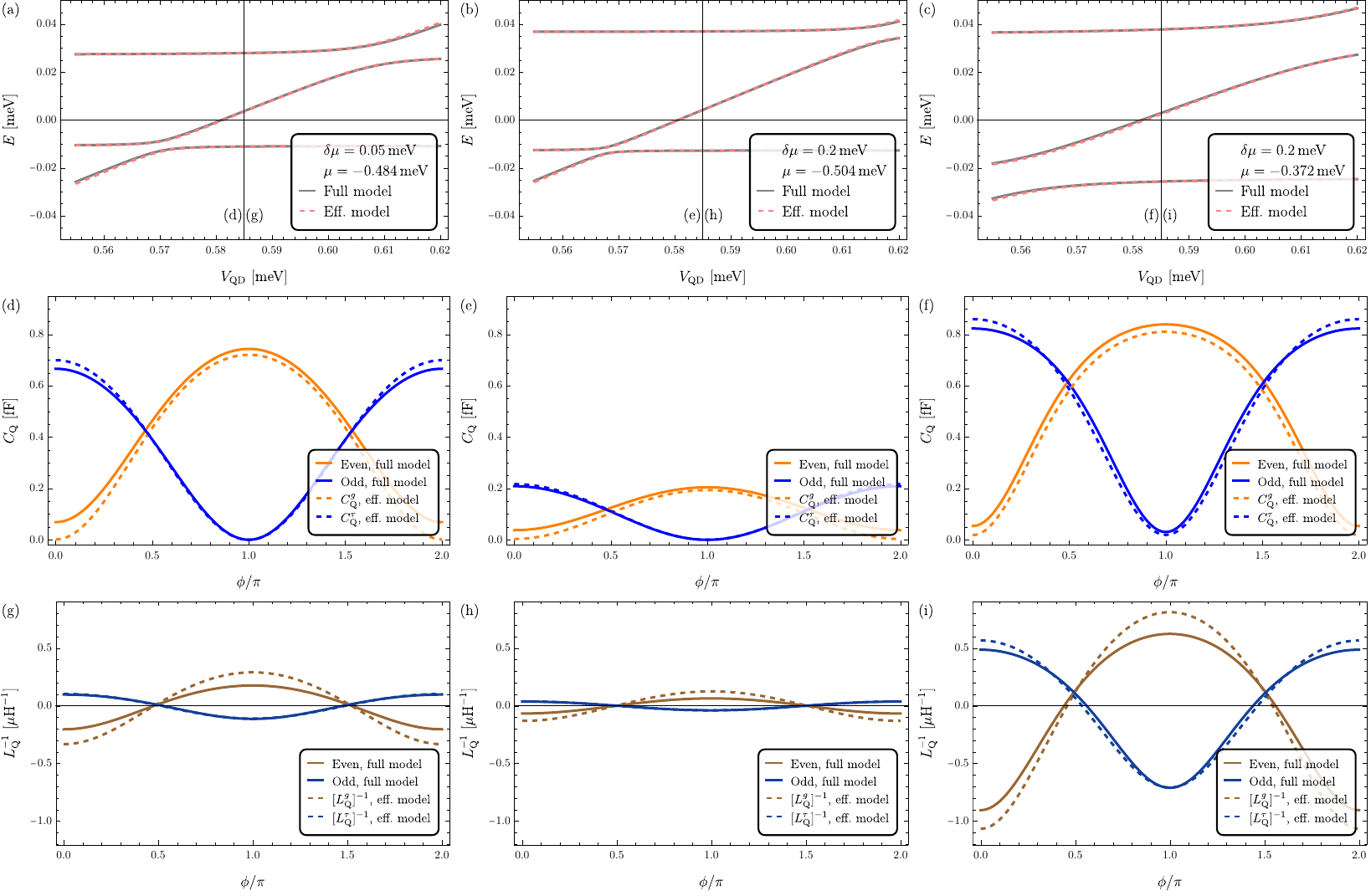}
    \caption{(a)--(c) Low-energy single-particle spectrum of the disordered normal interferometer as a function of $V_\m{QD}$ for a specific disorder realization, shown for different disorder strengths $\delta \mu$, and chemical potentials $\mu$.
    (d)--(f) Even and odd branches of the quantum capacitance $C_\m{Q}(\phi)$, and (g)--(i) quantum inductance $L_\m{Q}^{-1}(\phi)$, evaluated at $V_\m{QD} = 0.585\,\m{meV}$, indicated by the black lines in (a)--(c).
    The effective-model predictions are shown as dashed lines.
    In all cases, we observe Majorana-like $2\pi$ periodicity and an approximate $\pi$ phase shift between the even and odd branches.
    Disorder generally reduces the amplitudes of $C_\m{Q}$ and $L_\m{Q}^{-1}$.
    However, appropriate tuning of $\mu$ can restore the amplitudes, particularly for stronger disorder.
    The parameters are listed in Table~\ref{tab:normal_disorder} in Appendix~\ref{app:param}.
    }
    \label{fig:normal_results_disorder}
\end{figure*}

We investigate the effect of potential disorder on the normal interferometer.
The same parameters as in
    Figs.~\ref{fig:normal_results} (e) and (f)
    are used, and the potential disorder in the nanowire is modeled as $\mu_n = \mu + \delta \mu \, X_n$, where $X_n \sim \mathcal{N}(0,1)$ are independent and identically distributed random variables.
To obtain similar results to the disorder-free case, we fix $V_\m{QD} = 0.585\,\m{meV}$ and tune $\mu$.
In general, disorder reduces the magnitude of $C_\m{Q}$ and $L_\m{Q}^{-1}$.
However, the qualitative behavior remains similar to that of a clean system after appropriate tuning of $\mu$.

As an example, we consider a single disorder realization with fixed $X_n$, which captures the generic features of $C_\m{Q}$ and $L_\m{Q}^{-1}$.
For $\delta \mu = 0.05\,\m{meV}$, which is comparable to the level spacing of about $0.03\,\m{meV}$, the amplitudes of both quantities are reduced compared to the clean system
    [compare
        Figs.~\ref{fig:normal_results}(c) and (f)
        with 
        Figs.~\ref{fig:normal_results_disorder}(d) and (g)%
    ].
In this case, the quantum-dot level lies between the energy of the ground state and the first excited state of the nanowire
    [see Fig.~\ref{fig:normal_results_disorder}(a)].
Increasing the disorder strength to $\delta \mu = 0.2\,\m{meV}$ leads to a further suppression of the amplitudes
    [see Figs.~\ref{fig:normal_results_disorder}(e) and (h)].
However, by tuning the chemical potential to $\mu = -0.372\,\m{meV}$, where the quantum-dot level lies between the energies of the third and fourth excited states of the nanowire
    [see Fig.~\ref{fig:normal_results_disorder}(c)],
    the effective coupling strengths $g(\phi)$ and $\tau(\phi)$ are enhanced.
This results in a corresponding increase in the amplitudes of $C_\m{Q}$ and $L_\m{Q}^{-1}$
    [see Fig.~\ref{fig:normal_results_disorder}(f) and (i)].

\section{Parameter values}\label{app:param}

In this Appendix, we present all parameter values used in each figure (see Tables \ref{tab:Majorana} - \ref{tab:normal_disorder}). All energy values in the following tables are given in units of meV. The dashes in the tables mean that the corresponding parameter was not used in the calculation, while the asterisks indicate that the respective parameter runs over a finite interval, which is indicated in the figure. 

We use the lattice model phenomenologically. The lattice spacing $a$ only fixes the mapping to a continuum nanowire, $t = \hbar^2/(2m^* a^2)$, $\alpha = \alpha_R/(2 a)$, and $L = (N+1)\,a$.
Since our conclusions depend on the low-energy spectrum, gap sizes, level spacings, and dimensionless tunnel couplings, no particular value of $a$ is required. A material-specific choice of $a$ can be made a posteriori to match a desired effective mass $m^*$ and nanowire length $L$.

\begin{table}[h!]
\centering
\begin{tabular}{|c|c|c|c|c|c|c|c|c|c|}
\hline
Fig. &  $N$   &  $t$ &  $\mu$ &  $\Delta_0$  & $\alpha$ & $\Delta_\m{Z}$& $\lambda_\m{QD}$ & $\phi$ &$V_\m{QD}$ \\
\hline

\ref{fig:Majorana}(a) & 300 & 102 &  0 & 0.5 & 3.5  & * & - & - &-\\
\hline

\ref{fig:Majorana}(b) & 300 & 102 &  0 & 0.5 & 3.5  & 0.675 & - & -&-\\
\hline

\ref{fig:Majorana_results}(a) & 300 & 102 &  0 & 0.5 & 3.5 & 0.675 & 0.02 & 0 & *\\
\hline

\ref{fig:Majorana_results}(b) & 300 & 102 &  0 & 0.5 & 3.5 & 0.675 & 0.02 & $0, \pi$ & *\\
\hline

\ref{fig:Majorana_results}(c),(d) & 300 & 102 &  0 & 0.5 & 3.5& 0.675 & 0.02 &  * & 0.75\\
\hline

\ref{fig:Majorana_results}(e),(f) & 300 & 102 &  0 & 0.5 & 3.5 & 0.675  & 0.02& * & 0.82\\
\hline

\ref{fig:MBS_unbal}(a),(b) & 300 & 102 &  0 & 0.5 & 3.5& 0.675 & 0.015, 0.02 &  * & 0.745\\
\hline

\ref{fig:MBS_overlap}(a) & 175 & 102 &  0 & 0.5 & 3.5 & *  & -& - & -\\
\hline

\ref{fig:MBS_overlap}(b) & 175 & 102 &  0 & 0.5 & 3.5 & 1.1  & -& - & -\\
\hline

\ref{fig:MBS_overlap}(c),(d) & 175 & 102 &  0 & 0.5 & 3.5 & 1.1  & 0.02& * & 1.25\\
\hline

\end{tabular}
\caption{Parameters used to model the topological interferometer. In Fig.~\ref{fig:MBS_unbal}, we take $\lambda_\m{L} = 0.015$ and $\lambda_\m{R} = 0.02$}.
\label{tab:Majorana}
\end{table}

\begin{table}[]
\centering
\begin{tabular}{|c|c|c|c|c|c|c|c|c|c|c|c|}
\hline
Fig. &  $N_1$ &  $N_s$ &  $t$ & $t_s$ & $\mu$ &  $\mu_s$ &$\Delta_0$ &  $\alpha$ &  $\Delta_\m{Z}$ & $\phi$&$V_\m{QD}$ \\
\hline

\ref{fig:1ABS}(a) & 60 & 300 &  100 & 20 & 0 & 2 & 0.09 & 14.35 & * &-&-\\
\hline

\ref{fig:1ABS}(b) & 60 & 300 &  100 & 20 & 0 & 2 & 0.09 & 14.35 & 1.26 &-&-\\
\hline

\ref{fig:1ABS_results}(a)& 60 & 300 &  100 & 20 & 0 & 2 & 0.09 & 14.35 & 1.26 &0& *\\
\hline

\ref{fig:1ABS_results}(b) & 60 & 300 &  100 & 20 & 0 & 2 & 0.09 & 14.35 & 1.26 &$0, \frac{\pi}{2}, \pi$& *\\
\hline

\ref{fig:1ABS_results}(c),(d) & 60 & 300 &  100 & 20 & 0 & 2 & 0.09 & 14.35 & 1.26&*&1.425\\
\hline

\ref{fig:1ABS_results}(e),(f) & 60 & 300&  100 & 20 & 0 & 2 & 0.09 & 14.35 & 1.26 & *&1.38\\
\hline

\ref{fig:1ABS_results}(g),(h) & 60 & 300&  100 & 20 & 0 & 2 & 0.09 & 14.35 & 1.26 & *&1.48\\
\hline

\ref{fig:1ABS_not_mimic}(a)& 60 & 175&  100 & 20 & 0 & 2 & 0.25 & 14.35 & 1.435 & 0&*\\
\hline

\ref{fig:1ABS_not_mimic}(b) & 60 & 175&  100 & 20 & 0 & 2 & 0.25 & 14.35 & 1.435 & $0, \frac{\pi}{2}, \pi$&*\\
\hline

\ref{fig:1ABS_not_mimic}(c),(d) & 60 & 175&  100 & 20 & 0 & 2 & 0.25 & 14.35 & 1.435 & *&1.548\\
\hline

\ref{fig:1ABS_not_mimic}(e),(f) & 60 & 175&  100 & 20 & 0 & 2 & 0.25 & 14.35 & 1.435 & *&1.57\\
\hline

\ref{fig:1ABS_not_mimic}(g),(h) & 60 & 175&  100 & 20 & 0 & 2 & 0.25 & 14.35 & 1.435 & *&1.46\\
\hline

\end{tabular}
\caption{Parameters used to model the non-topological interferometer with one ABS. In this setup, $N_2=0$. In Fig.~\ref{fig:1ABS_results}, we take $\lambda_\m{L} = 0.01$ and $\lambda_\m{R} = 0.03$. In Fig.~\ref{fig:1ABS_not_mimic}, we take $\lambda_\m{QD} = 0.02$ everywhere.}
\label{tab:1ABS}
\end{table}

\begin{table}[]
\centering
\begin{tabular}{|c|c|c|c|c|c|c|c|c|c|c|c|c|c|}
\hline
Fig. &  $N_1$ &  $N_2$ &  $N_s$ & $t$ & $t_s$ & $\mu$ &  $\mu_s$ &$\Delta_0$ & $\alpha$&  $\Delta_\m{Z}$&  $\lambda_\m{QD}$ &$\phi$&$V_\m{QD}$ \\
\hline

\ref{fig:2ABS_pinned}(a) & 37 & 37 & 175 &  100 & 20 & 0 & 2 & 0.25 &14.35 & * & - & - &-\\
\hline

\ref{fig:2ABS_pinned}(b) & 37 & 37 & 175 &  100 & 20 & 0 & 2 & 0.25 &14.35 & 1.2 & - & - &-\\
\hline

\ref{fig:2ABS_pinned}(c),(d) & 37 & 37 & 175 &  100 & 20 & 0 & 2 & 0.25 &14.35 & 1.2 & 0.02 & *&1.261 \\
\hline

\ref{fig:2ABS_bulky}(a) & 60 & 42 & 230 &  100 & 20 & 0 & 2 & 0.09 &14.35 & * & - & - &-\\
\hline

\ref{fig:2ABS_bulky}(b) & 60 & 42 & 230 &  100 & 20 & 0 & 2 & 0.09 &14.35 & 1.08 & - & - &-\\
\hline

\ref{fig:2ABS_bulky}(c),(d) & 60 & 42 & 230 &  100 & 20 & 0 & 2 & 0.09 &14.35 & 1.08 & 0.02 & *&1.1 \\
\hline

\end{tabular}
\caption{Parameters used to model a non-topological nanowire with two ABS.
}
\label{tab:2ABS}
\end{table}

\begin{table}[]
\centering
\begin{tabular}{|c|c|c|c|c|c|c|c|c|}
\hline
Fig. &  $N$   &  $t$ &  $\mu$  & $\alpha$ & $\Delta_\m{Z}$& $\lambda_\m{QD}$ &$\phi$&$V_\m{QD}$ \\
\hline

\ref{fig:normal_results}(a) &  300&  102 &  $-0.48$  & 0 & 0.5& 0.02 &0&*\\
\hline

\ref{fig:normal_results}(b) &  300&  102 &  $-0.48$  & 0 & 0.5& 0.02 &$\pi/2$&*\\
\hline

\ref{fig:normal_results}(c),(d) &  300&  102 &  $-0.48$  & 0 & 0.5& 0.02 & *&0.575\\
\hline

\ref{fig:normal_results}(e),(f) &  300&  102 &  $-0.48$  & 0 & 0.5& 0.02 & *&0.585\\
\hline

\end{tabular}
\caption{Parameters used to model the normal interferometer.
}
\label{tab:normal}
\end{table}

\begin{table}[h!]
\centering
\begin{tabular}{|c|c|c|c|c|c|c|c|c|c|}
\hline
Fig.                                     & $N$ & $t$ & $\mu$    & $\delta \mu$
                                                                       & $\alpha$
                                                                           & $\Delta_\m{Z}$
                                                                                & $\lambda_\m{QD}$
                                                                                       &$\phi$
                                                                                                 &$V_\m{QD}$ \\\hline
\ref{fig:normal_results_disorder}(a)     & 300 & 102 & $-0.484$ & 0.05 & 0 & 0.5& 0.02 & $\pi/2$ & * \\\hline
\ref{fig:normal_results_disorder}(b)     & 300 & 102 & $-0.504$ & 0.2  & 0 & 0.5& 0.02 & $\pi/2$ & * \\\hline
\ref{fig:normal_results_disorder}(c)     & 300 & 102 & $-0.372$ & 0.2  & 0 & 0.5& 0.02 & $\pi/2$ & * \\\hline
\ref{fig:normal_results_disorder}(d),(g) & 300 & 102 & $-0.484$ & 0.05 & 0 & 0.5& 0.02 &     *   & 0.585 \\\hline
\ref{fig:normal_results_disorder}(e),(h) & 300 & 102 & $-0.504$ & 0.2  & 0 & 0.5& 0.02 &     *   & 0.585 \\\hline
\ref{fig:normal_results_disorder}(f),(i) & 300 & 102 & $-0.372$ & 0.2  & 0 & 0.5& 0.02 &     *   & 0.585 \\\hline

\end{tabular}
\caption{Parameters used to model the normal interferometer with potential disorder.
}
\label{tab:normal_disorder}
\end{table}

\end{document}